\documentclass[12pt]{article}
\usepackage{amsmath}
\usepackage{amssymb}
\usepackage{graphicx}
\usepackage{float}
\usepackage{indentfirst}
\usepackage{mathrsfs}
\usepackage{dsfont}

\usepackage{setspace}

\usepackage[labelfont=bf,labelsep=period]{caption}

\usepackage[numbers,sort&compress]{natbib}

\newtheorem{theorem}{Theorem}

\newtheorem{corollary}{Corollary}

\title{State-independent Uncertainty Relations and  Entanglement Detection}
\author
{Chen Qian, Jun-Li Li, and Cong-Feng Qiao$^{\ast}$\\ [0.2cm]
\normalsize{Department of Physics, University of Chinese Academy of Sciences,}\\
\normalsize{YuQuan Road 19A, Beijing 100049, China}\\[2pt]
\normalsize{Key Laboratory of Vacuum Physics, University of Chinese Academy of Sciences}\\[3mm]
\normalsize{$^\ast$ To whom correspondence should be addressed; E-mail: qiaocf@ucas.ac.cn.}
}

\date{}

\begin{document}
\baselineskip24pt \maketitle
\begin{abstract}\doublespacing
The uncertainty relation is one of the key ingredients of quantum theory. Despite the great efforts devoted to this subject, most of the variance-based uncertainty relations are state-dependent and suffering from the triviality problem of zero lower bounds. Here we develop a method to get uncertainty relations with state-independent lower bounds. The method works by exploring the eigenvalues of a Hermitian matrix composed by Bloch vectors of incompatible observables and is applicable for both pure and mixed states and for arbitrary number of $N$-dimensional observables. The uncertainty relation for incompatible observables can be explained by geometric relations related to the parallel postulate and the inequalities in Horn's conjecture on Hermitian matrix sum. Practical entanglement criteria are also presented based on the derived uncertainty relations.
\end{abstract}

\newpage

\section{Introduction}

The uncertainty relation is one of the distinguishing features of quantum theory and plays important roles in quantum information sciences \cite{PBusch, HHofmann, OGuhne, CAFuchs}. The original form, $p_1q_1\sim h$, was introduced by Heisenberg in explaining the non-simultaneous precision measurements of the position $q$ and the momentum $p$ of a microscopic particle, where $h$ is Planck constant and $p_1$ and $q_1$ are the precisions of measuring $p$ and $q$ \cite{heis}. Soon it was cast into the following form by Kennard \cite{Kennard}
\begin{eqnarray}
\Delta x\Delta p  \geq
\frac{\hbar}{2}\; . \label{heis}
\end{eqnarray}
Here $\Delta x$ and $\Delta p$ are the standard deviations in measuring the canonical observables of $x$ and $p$. The most well-known formulation, however, was that by Robertson \cite{Robertson}
\begin{eqnarray}\label{Robertson}
\Delta A^2 \Delta B^2\geq
\left|\frac{1}{2}\langle [A,B] \rangle\right|^2\; ,
\end{eqnarray}
where $\Delta X^2 = \langle X^2\rangle - \langle X\rangle^2$ means the variance (square of the standard deviation) of the observable (not just the canonical observables $x$ and $p$), and $[A,B]\equiv AB-BA$ is the commutator. Soon afterward Schr\"odinger presented an improvement, $ \Delta A^2\Delta B^2 \geq \left|\frac 12 \langle[A,B]\rangle \right|^2
+\left|\frac{1}{2} \langle\{A,B\}\rangle - \langle A\rangle\langle B\rangle\right|^2$ \cite{schrodinger},
with the anti-commutator defined as $\{A, B\} \equiv AB+BA$. These variance-based uncertainty relations have a common trait of state-dependent lower bound: The optimal lower bounds of the right hand sides may be trivially zero, which blurs the trade-offs between $\Delta A$ and $\Delta B$ for variant quantum states.

A recent work of Maccone and Pati's \cite{mp} presented new improvements to the uncertainty relation with a typical form of
\begin{eqnarray}
\Delta A^2 + \Delta B^2 & \geq &
\pm i\langle\psi|[A,B]|\psi\rangle+|\langle\psi|A \pm iB|\psi^\perp\rangle|^2 \; . \label{Maccone}
\end{eqnarray}
Here $|\psi^{\perp}\rangle$ is defined to be $\langle \psi|\psi^{\perp}\rangle =0$, and  the lower bound keeps on to be state-dependent. Since then, great efforts have been devoted to improve the lower bound of the variance-based uncertainty relation \cite{PU, F1, S1, S2, WUR, S3, Vur, WYI, MBP, Mprod, Uplow}. Of those new developments, the variance generally exists on both sides of the uncertainty relations, and thus the state-dependence remains. Moreover, those new uncertainty relations are mostly devoted to pure state and are hence not very suitable for mixed states \cite{mixed-un}. Generally, the infimum over all states for the right-hand sides of these uncertainty relations, e.g., see equations (\ref{Robertson}) and (\ref{Maccone}), will not give the real infimum for the left-hand sides. To get state-independent lower bounds, Bloch vector method was introduced in \cite{NewUR}, which may yield an exact uncertainty relation among arbitrary number of observables in principle \cite{TUR}. However, since the uncertainty relations obtained by means of Bloch vectors involve complicated functions of the variances of different observables \cite{NewUR, TUR}, the trade-offs among incompatible observables may not manifest explicitly. Numerical method is also helpful in analyzing the lower limits for the sum of observables' variances \cite{Neumerical-1}, e.g., variances of angular momentums \cite{Neumerical-2}. For the ever increasing number of uncertainty relations, the fundamental question remains open: How to get an explicit form of uncertainty relation with state-independent lower bound.

In this work, we present a method on how to derive the state-independent uncertainty relation for the sum of variances. The upper and lower bounds of the sum are obtained by exploring the eigenvalues of a Hermitian matrix composed of Bloch vectors of observables, which is applicable to both pure and mixed states and to arbitrary number of $N$-dimensional observables. In this sense, the quantum uncertainty relation stems from the geometric relation pertaining to the postulate in Euclidean geometry and the Horn's inequalities for the spectrum of Hermitian matrix sums \cite{Horn-conjecture} (the conjecture was proved around 2000 \cite{Horn-inequalities}). We also present a practical uncertainty-relation-based entanglement criterion for bipartite mixed states, which is shown to be superior to the Bloch representation criterion in detecting entanglement.

\section{The state-independent uncertainty relation}

An arbitrary quantum state (pure or mixed) may be represented by a density matrix. The density matrix $\rho$ is a positive semidefinite Hermitian matrix with trace one and may be expressed as \cite{N-vector}
\begin{eqnarray}
\rho = \frac{1}{N} \mathds{1} + \frac{1}{2} \sum_{\mu=1}^{N^2-1} r_{\mu}\lambda_{\mu} = \frac{1}{N} \mathds{1} + \frac{1}{2}\vec{r} \cdot \vec{\lambda}\; , \label{stat and obs}
\end{eqnarray}
where $\lambda_{\mu}$ are the $N^2-1$ SU($N$) generators with $\mathrm{Tr}[\lambda_{\mu}\lambda_{\nu}] = 2\delta_{\mu\nu}$, and $r_{\mu} = \mathrm{Tr}[\rho \lambda_{\mu}]$ are components of a $N^2-1$-dimensional real vector $\vec{r}$ called Bloch vector of the density matrix. The Bloch vector $\vec{r}$ subjects to a series of constraints to ensure the normalization and semipositivity of the density matrix \cite{N-bloch, N-Bloch-Positivity}.

In quantum mechanics, physical observables are represented by Hermitian matrices. Because adding (subtracting) a constant to (from) an observable does not change its variance, we can always treat the observable to be traceless and write $A = \sum_{\mu=1}^{N^2-1} a_{\mu}\lambda_{\mu} = \vec{a}\cdot \vec{\lambda}$, where $\vec{a}$ is called the Bloch vector of $A$. The variance of any observable $A$ in quantum state $\rho$ now can be written as \cite{NewUR}
\begin{eqnarray}
\Delta A^2 = \mathrm{Tr}[A^2\rho] -
\mathrm{Tr}[A\rho]^2 = \frac{2}{N} |\vec{a}|^2 + (\vec{a}*\vec{a}) \cdot \vec{r}
-(\vec{a}\cdot\vec{r}\,)^2\; . \label{Var-Bloch-N}
\end{eqnarray}
Here $(\vec{a}*\vec{a})_k = \sum_{\mu,\nu=1}^{N^2-1} a_{\mu} a_{\nu} d_{\mu\nu k}$ with $d_{\mu\nu k}$ being the symmetric structure constant of SU($N$) group. The variance of a physical observable now is expressed in terms of geometric relations between the Bloch vectors of the observable and the quantum state and varies with the states.

For $M$ observables $A_i=\vec{a}_i\cdot \vec{\lambda}$ in $N$-dimensional Hilbert space, we may construct a real symmetric matrix $\mathcal{A} = \sum_{i=1}^M \vec{a}_i \vec{a}_i^{\, \mathrm{T}}$. The Bloch vectors of $\{A_i\}$ span a space $\mathcal{S}_1 \equiv \mathrm{span}\{\vec{a}_i|i=1,\ldots, M\}$, where the whole $(N^2-1)$-dimensional Bloch vector space is constructed by $\mathcal{S}= \mathcal{S}_1 \cup \mathcal{S}_0$ with $\mathcal{S}_0 \equiv \overline{\mathcal{S}_1}$. The dimension $m$ of $\mathcal{S}_1$ lies in $1\leq m \leq \min\{M,N^2-1\}$. Then any Bloch vector can be decomposed accordingly as:
\begin{equation}
\vec{\alpha} = \sum_i^M \vec{a}_i * \vec{a}_i = \vec{\alpha}_1 + \vec{\alpha}_0\; , \; \vec{r} = \vec{r}_1 + \vec{r}_0 \; , \label{Space-Decom}
\end{equation}
where $\vec{\alpha}_1, \vec{r}_1 \in \mathcal{S}_1$ and $\vec{\alpha}_0, \vec{r}_0 \in \mathcal{S}_0$. We have the following theorem.
\begin{theorem}
For $M$ observables $A_i$, $i\in \{1,\ldots, M\}$, we have the following uncertainty relation
\begin{align}
\sum_{i=1}^M \Delta A_i^2 & \geq \frac{2}{N} \mathrm{Tr}[\mathcal{A}] + \mathcal{C}_0 - \mathcal{C}_{1} \; , \label{N-upper} \\
\sum_{i=1}^M \Delta A_i^2 & \leq \frac{2}{N} \mathrm{Tr}[\mathcal{A}] +  \mathcal{C}_0 - \mathcal{C}_{2} \; . \label{N-low}
\end{align}
Here $\mathcal{C}_0 = \frac{1}{4}\vec{\alpha_1}^{\mathrm{T}} \mathcal{A}^{-1} \vec{\alpha}_1$ is state independent, and
\begin{align}
\mathcal{C}_1 & = \max_{\theta \in [0,\pi/2]} \{(|\vec{r}\,| \sin\theta + \frac{1}{2} |\mathcal{A}^{-1} \vec{\alpha}_1|)^2\sigma_{1}(\mathcal{A}) + |\vec{\alpha}_0||\vec{r}\,|\cos\theta \} , \\
\mathcal{C}_2 & = \min_{\theta \in [0,\pi/2]} \{(|\vec{r}\,| \sin\theta - \frac{1}{2} |\mathcal{A}^{-1}\vec{\alpha}_1|)^2 \sigma_{m}(\mathcal{A}) - |\vec{\alpha}_0||\vec{r}\,| \cos\theta \} ,
\end{align}
where $\sigma_i(\cdot)$ are eigenvalues in descending order, $\mathcal{C}_{1}$ and $\mathcal{C}_{2}$ depend only on the norm of Bloch vector $|\vec{r}\,|$. \label{Theorem-N}
\end{theorem}

\noindent{\bf Proof:} According to equations (\ref{Var-Bloch-N}) and (\ref{Space-Decom}), we may write
\begin{align}
\sum_{i=1}^M \Delta A_i^2 = \frac{2}{N} \mathrm{Tr}[\mathcal{A}] + \frac{1}{4} \vec{\alpha_1}^{\mathrm{T}} \mathcal{A}^{-1} \vec{\alpha}_1 - (\vec{r}_1 - \frac{1}{2} \mathcal{A}^{-1} \vec{\alpha}_1)^{\mathrm{T}} \mathcal{A} (\vec{r}_1 - \frac{1}{2} \mathcal{A}^{-1} \vec{\alpha}_1) + \vec{\alpha}_0 \cdot \vec{r}_0 \; . \label{sum-Ai=}
\end{align}
Because $\mathcal{A} = \sum_{i=1}^M \vec{a}_i \vec{a}_i^{\, \mathrm{T}}$, it is invertible within $\mathcal{S}_1$. Equation (\ref{sum-Ai=}) has the lower bound
\begin{equation}
\sum_{i=1}^M \Delta A_i^2 \geq \frac{2}{N} \mathrm{Tr}[\mathcal{A}] +  \frac{1}{4} \vec{\alpha_1}^{\mathrm{T}} \mathcal{A}^{-1} \vec{\alpha}_1 - (|\vec{r}_1| +  \frac{1}{2} |\mathcal{A}^{-1} \vec{\alpha}_1|)^{2} \sigma_1(\mathcal{A}) - |\vec{\alpha}_0||\vec{r}_0| \; , \label{sum-Ai-lower}
\end{equation}
where $\sigma_1$ is the largest eigenvalue of $\mathcal{A}$. As $|\vec{r}\,|^2 = |\vec{r}_1|^2 + |\vec{r}_0|^2$, we have $|\vec{r}_1| = |\vec{r}\,|\sin\theta$ and $|\vec{r}_0| = |\vec{r}\,|\cos\theta$, $\theta \in [0,\pi/2]$, and therfore equation (\ref{sum-Ai-lower}) leads to equation (\ref{N-upper}). Equation (\ref{N-low}) is analogously obtained where $\sigma_m$ denotes the smallest eigenvalue of $\mathcal{A}$. Q.E.D.

As qubit systems have the most wide applications in quantum information sciences, we present several important corollaries of Theorem \ref{Theorem-N} for qubit. As $d_{\mu\nu k}=0$ for qubit, the variance in equation (\ref{Var-Bloch-N}) becomes $\Delta A^2  =  \vec{a}\cdot \vec{a}- (\vec{a} \cdot \vec{r}\,)^2$. We have the following Corollary
\begin{corollary}
For $M$ observables $A_i = \vec{a}_i \cdot \vec{\lambda}$ in qubit system, we have
\begin{eqnarray}
\sum_{i=1}^{M} \Delta A_{i}^2 & \geq & (1-|\vec{r}\,|^2) \sigma_{1} + \sigma_{2} + \sigma_{3}  \; , \label{l3} \\
\sum_{i=1}^{M} \Delta A_{i}^2 & \leq & \sigma_{1} + \sigma_{2} + (1-|\vec{r}\,|^2) \sigma_{3} \; . \label{l2}
\end{eqnarray}
Here $\sigma_i$ are eigenvalues of  $\mathcal{A} = \sum_{i=1}^M \vec{a}_i\vec{a}_i^{\,\mathrm{T}}$ with $\sigma_1\geq \sigma_2 \geq \sigma_3\geq 0$. \label{Theorem-singular}
\end{corollary}

\noindent{\bf Proof:} As $\vec{\alpha} = 0$ and $\Delta A^2  =  \vec{a}\cdot \vec{a}- (\vec{a} \cdot \vec{r}\,)^2$, it is easy to get the following result:
\begin{eqnarray}
\sum_{i=1}^{M} \Delta A_{i}^2 =  \mathrm{Tr}[\mathcal{A}] - \vec{r}^{\,\mathrm{T}} \mathcal{A} \, \vec{r} \; . \label{Var-norm-expression}
\end{eqnarray}
Because $\mathcal{A}$ is a positive semi-definite real symmetric matrix with eigenvalues $\{\sigma_1, \sigma_2, \sigma_3\}$, we have $|\vec{r}\,|^2 \sigma_3 \leq \vec{r}^{\,\mathrm{T}} \mathcal{A} \, \vec{r} \leq |\vec{r}\,|^2\sigma_1$
which directly leads to equations (\ref{l3}, \ref{l2}).
Q.E.D.

Corollary \ref{Theorem-singular} gives both the upper and lower bounds for the sum of the variances of $A_i$, which rely only on trace norm of the density matrix, i.e., $|\vec{r}\,|^2$. For the special case of pure qubit state where $|\vec{r}\,|=1$, we have
\begin{equation}
\sigma_2 + \sigma_3 \leq \sum_{i=1}^{M} \Delta A_{i}^2  \leq \sigma_1 + \sigma_2 \; .
\end{equation}
It is noticed that the inequalities in Theorem \ref{Theorem-singular} actually arise from the Horn's inequalities for the sum of Hermitian matrices, which will be clear with the following Corollary:
\begin{corollary}
For two independent observables $A_1$ and $A_2$ in qubit system, i.e., the two observables are not proportional $A_1 \neq \kappa A_2$, where $ \Delta A_1^2 \geq c_1$ and $\Delta A_2^2 \geq c_2$ with $c_{1}$ and $c_2$ being dependent only on the Bloch vector norm of the state, there exists the following
\begin{equation}
\Delta A_1^2  +  \Delta A_2^2 > c_1+c_2 \; .
\end{equation}
That is, the lower bound of the sum of their variances are greater than the sum of their variances' lower bounds for all the states with the same Bloch vector norm.
\end{corollary}

\noindent{\bf Proof:} According to Theorem \ref{Theorem-singular}, we have
\begin{align}
\Delta A_1^2 = |\vec{a}_1|^2 - \vec{r}^{\mathrm{T}} \mathcal{A}_1 \vec{r} \geq |\vec{a}_1|^2 - |\vec{r}\,|^2\sigma_1(\mathcal{A}_1) \equiv c_1\; , \\
\Delta A_2^2 = |\vec{a}_2|^2 - \vec{r}^{\mathrm{T}} \mathcal{A}_2 \vec{r} \geq |\vec{a}_2|^2 - |\vec{r}\,|^2\sigma_1(\mathcal{A}_2) \equiv c_2 \; ,
\end{align}
where $\vec{a}_i$ are the Bloch vectors of $A_i$; $\mathcal{A}_i = \vec{a}_i \vec{a}_i^{\,\mathrm{T}}$, $i=1,2$ are real symmetric (Hermitian) matrices; and $\sigma_1(\cdot)$ means the largest eigenvalue of a matrix. Meanwhile, the sum of the two variances is
\begin{equation}
\Delta A_{1}^2 + \Delta A_2^2 = \vec{a}_1 \cdot \vec{a}_1 + \vec{a}_2 \cdot \vec{a}_2- \vec{r}^{\,\mathrm{T}}( \mathcal{A}_1 + \mathcal{A}_2 ) \vec{r} \; . \label{sum-lower-bound}
\end{equation}
The lower bound of equation (\ref{sum-lower-bound}) is  $|\vec{a}_1|^2 + |\vec{a}_2|^2 - |\vec{r}\,|^2 \sigma_1(\mathcal{A}_1 + \mathcal{A}_2)$. However, the Horn's inequalities \cite{Horn-inequalities} tell that $\sigma_1(\mathcal{A}_1 + \mathcal{A}_2) < \sigma_1(\mathcal{A}_1) + \sigma_1(\mathcal{A}_2)$ for the present configuration of $\mathcal{A}_1$ and $\mathcal{A}_2$, i.e. $\vec{a}_1 \neq \kappa \vec{a}_2$.
Q.E.D.

Two physical quantities $A_1$ and $A_2$ may be regarded as linearly independent in the sense that $xA_1+yA_2 =0 \Leftrightarrow x=y=0$. In classical probability theory, the linear independence leads to the following: If there are probability distributions for $A_1$ and $A_2$ where $\Delta A_{1,2}^2$ could reach the values $c_{1,2}$ respectively, then there always exist the joint probability distribution that makes $\Delta A_1^2 + \Delta A_2^2 = c_1+c_2$, e.g., multiplying the two probability distributions will simply do the job. However, the quantum theory predicts differently: There does not exist the state (joint probability distribution for $A_1$ and $A_2$ in the statistical interpretation in quantum mechanics) where the variances of $A_1$ and $A_2$ could reach the individual minimum values of $c_1$ and $c_2$ simultaneously, i.e., $\Delta A_1^2 + \Delta A_2^2$ cannot reach $c_1+c_2$ due to the Horn's inequalities on the matrix sum in equation (\ref{sum-lower-bound}).

Similar situation as that of equation (\ref{sum-lower-bound}) happens in $N$-dimensional systems, i.e.
\begin{equation}
\Delta A_1^2 + \Delta A_2^2 = \frac{2}{N}\left(|\vec{a}_1|^2 + |\vec{a}_2|^2 \right) + (\vec{a}_1' + \vec{a}_2')\cdot \vec{r} - \vec{r}^{\, \mathrm{T}}(\mathcal{A}_1 + \mathcal{A}_2) \vec{r} \; ,
\end{equation}
where $\vec{a}_i' = \vec{a}_i*\vec{a}_i$ and $\mathcal{A}_i = \vec{a}_i \vec{a}_i^{\mathrm{T}}$. If $|\vec{a}_i'| \gg \sigma_1(\mathcal{A}_{i})$, then $\Delta A_1^2 + \Delta A_2^2 >c_1+c_2$ because $\vec{r}$ cannot be anti-parallel (or parallel) to two non-parallel vectors $\vec{a}_1'$ and $\vec{a}_2'$ simultaneously, which is originated from the parallel postulate of Euclidean geometry. While if $|\vec{a}_i'| \ll \sigma_1(\mathcal{A}_{i})$, $\Delta A_1^2 + \Delta A_2^2 >c_1+c_2$ satisfies, due to the Horn's inequalities for matrix sum (see equation (\ref{sum-lower-bound})). The actual case of $\frac{|\vec{a}_i'|}{\sigma_1(\mathcal{A}_i)} \in [0, \left(\frac{2(N-1)}{N}\right)^{\frac{1}{2}}]$ \cite{NewUR} may be more complex because of the possible interferences between terms $\vec{a}*\vec{a}$ and  $\vec{a} \vec{a}^{\,\mathrm{T}}$. However for a complete set of orthogonal observables $\{A_i\}$, where $\mathrm{Tr}[A_iA_j] = 2|\vec{a}\,|^2 \delta_{ij}$, $i,j \in \{1,\ldots, N^2-1\}$, there exists a concise result for the variance-based uncertainty relation. We express a complete set of orthogonal observables $\{A_i\}$ in the form of
\begin{equation}
A_{i} = |\vec{a}| \sum_{j =1}^{N^2-1} O_{ij} \lambda_{j} \; , \label{orth-A}
\end{equation}
where $O\in$ SO($N^2-1$), and we have
\begin{corollary}
For the complete set of orthogonal observables $\{A_i\}$, there exists the following relation for the state-dependent variances
\begin{eqnarray}
\sum_{i=1}^{N^2-1} \Delta A_i ^2 = |\vec{a}|^2 \left(\frac{2(N^2-1)}{N} - |\vec{r}\,|^2 \right) \geq 2|\vec{a}|^2 (N-1) \; .
\end{eqnarray}
Here $\vec{r}$ is the Bloch vector of the quantum state. \label{Theorem-ortho-complete}
\end{corollary}

\noindent{\bf Proof:} Taking equation (\ref{orth-A}) into equation (\ref{Var-Bloch-N}) we have
\begin{equation}
\Delta A_i^2 = \frac{2}{N} |\vec{a}|^2 + |\vec{a}|^2 \sum_{\mu,\nu, k} O_{i\mu}O_{i\nu}d_{\mu\nu k} r_k - |\vec{a}|^2(\sum_{\mu=1}^{N^2-1} O_{i\mu}r_{\mu})^2 \; .
\end{equation}
Summing over $i$, we have
\begin{align}
\sum_{i=1}^{N^2-1} \Delta A_i^2 & = |\vec{a}|^2 \frac{2(N^2-1)}{N} + |\vec{a}|^2 \sum_{\mu,k=1}^{N^2-1} d_{\mu\mu k} r_{k} - |\vec{a}|^2 |\vec{r}\,|^2 \nonumber \\
& = |\vec{a}|^2 \left(\frac{2(N^2-1)}{N} - |\vec{r}\,|^2 \right) \; ,
\end{align}
where the condition $\sum_{\mu=1}^{N^2-1} d_{\mu\mu k} = 0$, $\forall k \in \{1,\ldots, N^2-1\}$ is employed. As the Bloch vectors $|\vec{r}\,|^2 \leq \frac{2(N-1)}{N}$, the theorem is then sound. Q.E.D.

Corollary \ref{Theorem-ortho-complete} states that when a complete set of orthogonal observables is considered, the sum of their variances appears to be an identity. For Pauli matrices in SU(2), Corollary \ref{Theorem-ortho-complete} reduces to $\Delta \sigma_1^2 + \Delta \sigma_2^2 + \Delta \sigma_3^2 = 3- |\vec{r}\,|^2$ which agrees with the result of \cite{NewUR}.

\section{The detection of entanglement via uncertainty relations}

Uncertainty relations can also be used to characterize quantum entanglement \cite{Quant}. We consider the following $N\times N$ quantum state in Bloch representation \cite{Separability-Horn}
\begin{align}
\rho_{AB} & =  \frac{1}{N^2}  \mathds{1} \otimes \mathds{1} + \frac{1}{2N} \vec{r}\cdot \vec{\lambda}\otimes \mathds{1} + \frac{1}{2N} \mathds{1} \otimes \vec{s} \cdot \vec{\lambda} + \frac{1}{4} \sum_{\mu=1}^{N^2-1} \sum_{\nu=1}^{N^2-1} \mathcal{T}_{\mu\nu} \, \lambda_{\mu} \otimes \lambda_{\nu} \; ,  \label{rho-Bloch}
\end{align}
where $r_{\mu} = \mathrm{Tr}[\rho_{AB}(\lambda_{\mu} \otimes \mathds{1})]$, $s_{\nu} = \mathrm{Tr}[\rho_{AB} (\mathds{1} \otimes \lambda_{\nu})]$, and $\mathcal{T}_{\mu\nu} = \mathrm{Tr}[\rho_{AB} (\lambda_{\mu} \otimes \lambda_{\nu})]$ is called the correlation matrix. The reduce density matrices are $\rho_A = \mathrm{Tr}_B[\rho_{AB}] = \frac{1}{N}\mathds{1} + \frac{1}{2}\vec{r} \cdot \vec{\lambda}$,  $\rho_B = \mathrm{Tr}_A[\rho_{AB}] = \frac{1}{N}\mathds{1} + \frac{1}{2}\vec{s} \cdot \vec{\lambda}$, and the quantum state $\rho_{AB}$ is separable when
\begin{equation}
\vec{r} = \sum_k p_k \vec{r}_k\; , \vec{s} = \sum_k p_k \vec{s}_k\; , \; \mathcal{T} = \sum_{k} p_k \vec{r}_k \vec{s}_k^{\, \mathrm{T}}\; .
\end{equation}
Here, $\rho_{AB} = \sum_k p_k \rho^{(A)}_k \otimes \rho^{(B)}_{k}$, $\{p_i\}$ is probability distribution, and $\vec{r}_k$ and $\vec{s}_k$ denote the Bloch vectors of $\rho_k^{(A)}$ and $\rho_k^{(B)}$ respectively. We call a set of local observables $M_i = A_i\otimes \mathds{1} + \mathds{1} \otimes B_i$ to be complete and orthonormal if $\mathrm{Tr}[A_iA_j] = \mathrm{Tr}[B_iB_j] = 2\delta_{ij}$, $\forall i,j \in \{1,\ldots, N^2-1\}$, and the following Corollary exists:
\begin{corollary}
If an $N\times N$ state $\rho_{AB}$ is separable, then the following relation exists for arbitrary complete orthonormal local observables $\{M_i=A_i\otimes \mathds{1} + \mathds{1} \otimes B_i\}$
\begin{align}
\sum_{i=1}^{N^2-1} \Delta M_i^2  &  \geq 4(N-1) \; . \label{uncertainty-ent}
\end{align}
Equation (\ref{uncertainty-ent}) directly tells that if $\rho_{AB}$ is separable, then
\begin{align}
||\mathcal{T}||_{\mathrm{KF}} & \leq \frac{2(N-1)}{N} - \frac{1}{2} \left( |\vec{r}\,| - |\vec{s}\,| \right)^2 \; , \label{uncertainty-BR}
\end{align}
where $||\mathcal{T}||_{\mathrm{KF}} \equiv \sum_i \sigma_{i}(\mathcal{T})$ is the Ky Fan norm of a matrix, $\vec{r}$ and $\vec{s}$ are the Bloch vectors of the reduce density matrices of particles $A$ and $B$. \label{Coro-separable}
\end{corollary}

\noindent{\bf Proof:} Taking equation (\ref{rho-Bloch}) into $\Delta M_i^2 = \mathrm{Tr}[M_i^2 \rho_{AB}] - \mathrm{Tr}[M_i\rho_{AB}]^2$, we have
\begin{align}
\sum_{i=1}^{N^2-1} \Delta M_i^2 & = \frac{4}{N}(N^2-1) + 2 \sum_i \vec{a}_i^{\,\mathrm{T}} \mathcal{T} \vec{b}_i - \sum_i  (\vec{r} \cdot \vec{a}_i + \vec{s} \cdot \vec{b}_i)^2\nonumber \\ & \leq \frac{4}{N}(N^2-1) + 2 \sum_i \vec{a}_i^{\,\mathrm{T}} \mathcal{T} \vec{b}_i - \left( |\vec{r}\,| - |\vec{s}\,| \right)^2 \; . \label{DeltaMi-NS}
\end{align}
While taking equation (\ref{rho-Bloch}) into $\Delta M_i^2 = \mathrm{Tr}[M_i^2\rho_{AB}] - \mathrm{Tr}[M_i\rho_{AB}]^2$ with $\mathcal{T} = \sum_{k} p_k \vec{r}_k \vec{s}_k^{\mathrm{T}}$, we have
\begin{align}
\sum_{i=1}^{N^2-1} \Delta M_i^2 & = \frac{4(N^2-1)}{N} + 2 \sum_{i,k} p_k r_{ki}s_{ki} - \sum_i \left(\sum_{ k}p_kr_{ki}+ p_ks_{ki} \right)^2 \nonumber \\
& \geq \frac{4(N^2-1)}{N} - \sum_{k}p_k (|\vec{r}_k|^2+|\vec{s}_k|^2)  \nonumber \\
& \geq 4(N-1) \; . \label{DeltaMi-S}
\end{align}
Here, $r_{ki}=\vec{r}_k\cdot \vec{a}_i$, $s_{ki}=\vec{s}_k\cdot \vec{b}_i$, and $|\vec{r}\,|^2, |\vec{s}\,|^2 \leq \frac{2(N-1)}{N}$; the relation $2\sum_{k} p_k r_{ki}s_{ki} = \sum_{k} p_k \left[(r_{ki}+s_{ki})^2 - r_{ki}^2 - s_{ki}^2 \right]$
is used. Then equations (\ref{DeltaMi-NS}, \ref{DeltaMi-S}) give
\begin{equation}
\sum_{i=1}^{N^2-1} \vec{a}_i^{\,\mathrm{T}} \mathcal{T} \vec{b}_i \geq - \frac{2(N-1)}{N} +\frac{1}{2} \left( |\vec{r}\,| - |\vec{s}\,| \right)^2\; , \label{pre-result}
\end{equation}
which is satisfied by all possible bases $\vec{a}_i$ and $\vec{b}_i$. By choosing $\vec{a}_i = \vec{u}_i$ and $\vec{b}_i = -\vec{v}_i$, we have
\begin{equation}
||\mathcal{T}||_{\mathrm{KF}} \leq  \frac{2(N-1)}{N} - \frac{1}{2} \left( |\vec{r}\,| - |\vec{s}\,| \right)^2 \; .
\end{equation}
Here $\vec{u}_i$ and $\vec{v}_i$ are the left and right singular vectors of $\mathcal{T}$. Q.E.D.

Corollary \ref{Coro-separable} represents an uncertainty-relation-based entanglement criterion for bipartite mixed states. Equation (\ref{uncertainty-BR}) provides a  better upper bound than Theorem 1 of Ref. \cite{Separable-BR}. When the subsystems of an $N\times N$ quantum state are completely mixed, i.e. $|\vec{r}\,|=|\vec{s}\,|=0$, the computable cross-norm or realignment (CCNR) criterion \cite{norm-Rudolph, norm-realign}, Bloch representation criterion \cite{Separable-BR}, the covariance matrix criterion \cite{covariance-matrix} and the local uncertainty relation criterion of equation (\ref{uncertainty-BR}) all converge to the same relation: $||\mathcal{T}||_{\mathrm{KF}} \leq  \frac{2(N-1)}{N}$. Most importantly, our method also provides a way to construct the optimal observable set $\{M_i\}$ to detect the entanglement, which is generally a difficult task for the uncertainty-relation-based entanglement criteria. That is, the Bloch vectors should be properly chosen according to the left and right singular vectors of the correlation matrix $\mathcal{T}$.

\section{Conclusion}

In this work we have proposed a state-independent variance-based uncertainty relation by virtue of the observables' Bloch vectors. By exploring the eigenvalues of the Hermitian matrix composed of the Bloch vectors, the upper and lower bounds for the sum of variances are obtained. It is found that the incompatibility of observables may be attributed to geometric relations related to the parallel postulate of Euclidean geometry and the Horn's conjecture on the Hermitian matrix sum, which provides an alternative interpretation for the variance-based uncertainty relation. Also, our method leads to a practical entanglement criterion for bipartite mixed states. Considering the important roles it plays in the separability problem \cite{Separability-Horn, Separable-decomp}, we believe the Bloch representation can be as a useful tool for further quantitative study and deeper understanding of the fundamental concepts in quantum theory, e.g. the uncertainty relation, quantum entanglement, and quantum steering \cite{Steering}.

\section*{Acknowledgements}
This work was supported in part by the Ministry of Science and Technology of the People's Republic of China(2015CB856703); by the Strategic Priority Research Program of the Chinese Academy of Sciences, Grant No.XDB23030100; and by the National Natural Science Foundation of China(NSFC) under the Grants 11375200 and 11635009.


\end{document}